\documentclass[
reprint, amsmath,amssymb,
 aps, prl, twocolumn
]{revtex4-2}

\usepackage{graphicx}
\usepackage{dcolumn}
\usepackage{bm}

\begin{document}

\title{\textbf{Shake-off in XFEL heated solid density plasma}}%

\author{G. O. Williams}
 \email{Contact author: gareth.williams@tecnico.ulisboa.pt}
 \affiliation{GoLP/Instituto de Plasmas e Fus\~ao Nuclear, Instituto Superior T\'ecnico, Universidade de Lisboa, 1049-001 Lisboa, Portugal}

\author{L. Ansia}%
\affiliation{GoLP/Instituto de Plasmas e Fus\~ao Nuclear, Instituto Superior T\'ecnico, Universidade de Lisboa, 1049-001 Lisboa, Portugal}
\altaffiliation{Instituto de Fusi\'on Nuclear, Universidad Polit\'ecnica de Madrid, Jos\'e Guti\'errez Abascal 2, 28006 Madrid, Spain}

\author{M. Makita}
\affiliation{European XFEL, Holzkoppel 4, 22869 Schenefeld, Germany}

\author{P. Estrela}
\affiliation{GoLP/Instituto de Plasmas e Fus\~ao Nuclear, Instituto Superior T\'ecnico,
Universidade de Lisboa, 1049-001 Lisboa, Portugal}

\author{M. Hussain}
\affiliation{GoLP/Instituto de Plasmas e Fus\~ao Nuclear, Instituto Superior T\'ecnico,
Universidade de Lisboa, 1049-001 Lisboa, Portugal}

\author{T. R. Preston}
\affiliation{European XFEL, Holzkoppel 4, 22869 Schenefeld, Germany}

\author{J. Chalupsk\'{y}}
\affiliation{Department of Radiation and Chemical Physics, Institute of Physics, Czech Academy of Sciences, Na Slovance 1999/2, 182 21 Prague, Czech Republic}

\author{V. Hajkova}
\affiliation{Department of Radiation and Chemical Physics, Institute of Physics, Czech Academy of Sciences, Na Slovance 1999/2, 182 21 Prague, Czech Republic}

\author{T. Burian}
\affiliation{Department of Radiation and Chemical Physics, Institute of Physics, Czech Academy of Sciences, Na Slovance 1999/2, 182 21 Prague, Czech Republic}

\author{M. Nakatsutsumi}
\affiliation{European XFEL, Holzkoppel 4, 22869 Schenefeld, Germany}

\author{J. Kaa}
\affiliation{European XFEL, Holzkoppel 4, 22869 Schenefeld, Germany}

\author{Z. Konopkova}
\affiliation{European XFEL, Holzkoppel 4, 22869 Schenefeld, Germany}

\author{N. Kujala}
\affiliation{European XFEL, Holzkoppel 4, 22869 Schenefeld, Germany}

\author{K. Appel}
\affiliation{European XFEL, Holzkoppel 4, 22869 Schenefeld, Germany}

\author{S. G{\"o}de}
\affiliation{European XFEL, Holzkoppel 4, 22869 Schenefeld, Germany}

\author{V. Cerantola}
\affiliation{European XFEL, Holzkoppel 4, 22869 Schenefeld, Germany}

\author{L. Wollenweber}
\affiliation{European XFEL, Holzkoppel 4, 22869 Schenefeld, Germany}

\author{E. Brambrink}
\affiliation{European XFEL, Holzkoppel 4, 22869 Schenefeld, Germany}

\author{C. Baehtz}
\affiliation{Helmholtz-Zentrum Dresden-Rossendorf, Bautzner Landstra{\ss}e 400, Dresden, 01328, Germany}

\author{J-P. Schwinkendorf}
\affiliation{European XFEL, Holzkoppel 4, 22869 Schenefeld, Germany}

\author{V. Vozda}
\affiliation{Department of Radiation and Chemical Physics, Institute of Physics, Czech Academy of Sciences, Na Slovance 1999/2, 182 21 Prague, Czech Republic}

\author{L. Juha}
\affiliation{Department of Radiation and Chemical Physics, Institute of Physics, Czech Academy of Sciences, Na Slovance 1999/2, 182 21 Prague, Czech Republic}

\author{H.-K. Chung}
\affiliation{Korea Institute of Fusion Energy, Yuseong-gu, Daejeon, 34133, Republic of Korea}

\author{P. Vagovic}
\affiliation{Center for Free-Electron Laser Science CFEL, Deutsches ElektronenSynchrotron DESY, 22607 Hamburg, Germany}

\author{H. Scott}
\affiliation{Lawrence Livermore National Laboratory, Livermore, California 94550, USA}

\author{P. Velarde}
\affiliation{Instituto de Fusi\'on Nuclear, Universidad Polit\'ecnica de Madrid, Jos\'e Guti\'errez Abascal 2, 28006 Madrid, Spain}

\author{U. Zastrau}
\affiliation{European XFEL, Holzkoppel 4, 22869 Schenefeld, Germany}

\author{M. Fajardo}
\affiliation{GoLP/Instituto de Plasmas e Fus\~ao Nuclear, Instituto Superior T\'ecnico,
Universidade de Lisboa, 1049-001 Lisboa, Portugal}

\date{\today}

\begin{abstract}
In atoms undergoing ionisation, an abrupt re-arrangement of free and bound electrons can lead to the  ejection of another bound electron (shake-off). The spectroscopic signatures of shake-off have been predicted and observed in atoms and solids. Here, we present the first observation of this process in a solid-density plasma heated by an x-ray free electron laser. The results show that shake-off of L-shell electrons persists up to temperatures of 10 eV at solid density, and follow the probability predicted for solids. This work shows that shake-off should be included in plasma models for the correct interpretation of emission spectra.
\end{abstract}
\maketitle

X-ray free electron lasers (XFELs) have driven a series of new discoveries in high energy density physics \cite{Vinko2012,Fletcher2015,LasoGarcia2024}. The capability to precisely deposit the x-ray energy in femtoseconds has allowed the creation of gradient-free defined plasmas \cite{Levy2015}. Bright coherent XFEL pulses have allowed new plasma diagnostic capabilities using ultrabright scattering of laser produced plasmas \cite{Fletcher2015}. XFEL's can now heat solids to above 100 eV, while emission spectra from radiative recombination following the XFEL induced core vacancies can reveal ion distributions during the XFEL pulse \cite{Vinko2012},  ionisation potential depression \cite{Ciricosta2012}, and collisional cross sections at solid density \cite{Berg2018}. Experiments largely rely on the measurement of K-alpha satellite lines, which shift to higher photon energies for higher ionisation stages. Following core electron photoionisation by the XFEL, further ionisation was generally thought to occur through collisional pathways and Auger emission (auto-ionisation). Up to now, collisional ionisation has been considered the dominant ionisation process and source of K-alpha satellite peaks in XFEL heated solid plasmas. However, another ionisation process that has been overlooked but can result in satellite emission lines is the abrupt emission of a secondary electron to the continuum during a primary ionisation process, called shake-off. When an atom undergoes an inner-shell ionisation the ion potential abruptly changes as viewed by the remaining bound electrons. This sudden change can cause a secondary electron excitation (shake-up) or emission (shake-off). In solids or atoms where there are few or no states available for shake-up to occur, shake-off is the dominant pathway. Unlike Auger emission, shake-off does not occur through a bound shell recombination to provide the energy source for secondary ionisation: shake-off shares the energy of the ionising particle between the ejected electrons, resulting in a doubly ionised atom. This process was first observed experimentally in experiments of ionisation during beta decay \cite{Snell1957}. Later experiments observed similar secondary ionisation during photoionisation \cite{Carlson1965}, which drove theories to describe secondary electron emission during photoionisation \cite{Thomas1984, Carlson1973}. The theory and calculation method was advanced by \r{A}rberg \cite{Aaberg1976}. Following photoionisation, this process leaves an atom with two vacancies: one due to the initial photoionisation, and another due to the 'shaking-off' of another electron in the same atom due to the sudden change in potential. When radiative recombination occurs in this atom, the spectroscopic signature will show emission of this higher charge state in satellite peaks at higher energies. Modern XFELs provide an ideal platform to study the magnitude of shake-off events. Conditions are met when the photon has sufficient energy to create the first \textit{and} second vacancy which can in principle be any electron, but the probability generally increases for outer electrons with lower ionisation energy \cite{Mukoyama1987}. In this study we focus on L-shell shake-off following K-shell photoionisation which requires $E_{X} = E_K + E_L$, where $E_{X}$ is the XFEL photon energy, $E_K$ is the K-shell ionisation energy, and $E_L$ is the L-shell ionisation energy. Shake-off processses have been studied as a function of photon energy in cold solids, and were shown to follow analytical predictions \cite{Oura2003}.  The question as to whether this process persists in dense plasmas has so far remained unanswered, yet is crucial for the correct interpretation of measured emission spectra and our understanding of dense plasmas.

Here, we use an XFEL to both heat and induce fluorescence in solid titanium, and show for the first time that shake-off processes occur with a very similar probabliltiy to that of cold solids. These results show that the shake-off probability is unchanged from the solid to the plasma state at temperatures of up to 10 eV. Shake-off leads to ionisation and satellites in the emission spectra that are not created through standard photo or collisional processes.

The experiment was conducted at the high energy density HED end station at the European XFEL \cite{Zastrau2021}. X-ray pulses were focussed using onto titanium foils of 1 $\mu$m in thickness using compound refractive beryllium lenses. Emission spectra were recorded using two x-ray spectrometers \cite{Preston2020}. The intensity on target was measured using upstream and downstream energy monitors and imprint analysis of the damaged areas \cite{Chalupsky2013}. Focussed XFEL effective spot diameters were 10 $\mu$m and peak intensities of ~$10^{16}$ W cm$^{-2}$ were reached in the central region. These intensities result in peak electron temperatures of $\sim$10 eV. The samples were heated with two distinct XFEL photon energies, 5.1 and 6 keV, to distinguish between the results in which L-shell shake-off is not possible and those in which it is possible. The XFEL photon energy spectrum was monitored with the HIREX2 spectrometer \cite{Kujala2020}.  Pulse energies were controlled with attenuators to record spectra at a range of intensities to investigate the effect of varying plasma temperature. The XFEL pulse duration was estimated to be 25 fs at full width at half maximum, ensuring the emission spectrum is recorded at solid density.

\begin{figure}
\includegraphics[scale=0.42]{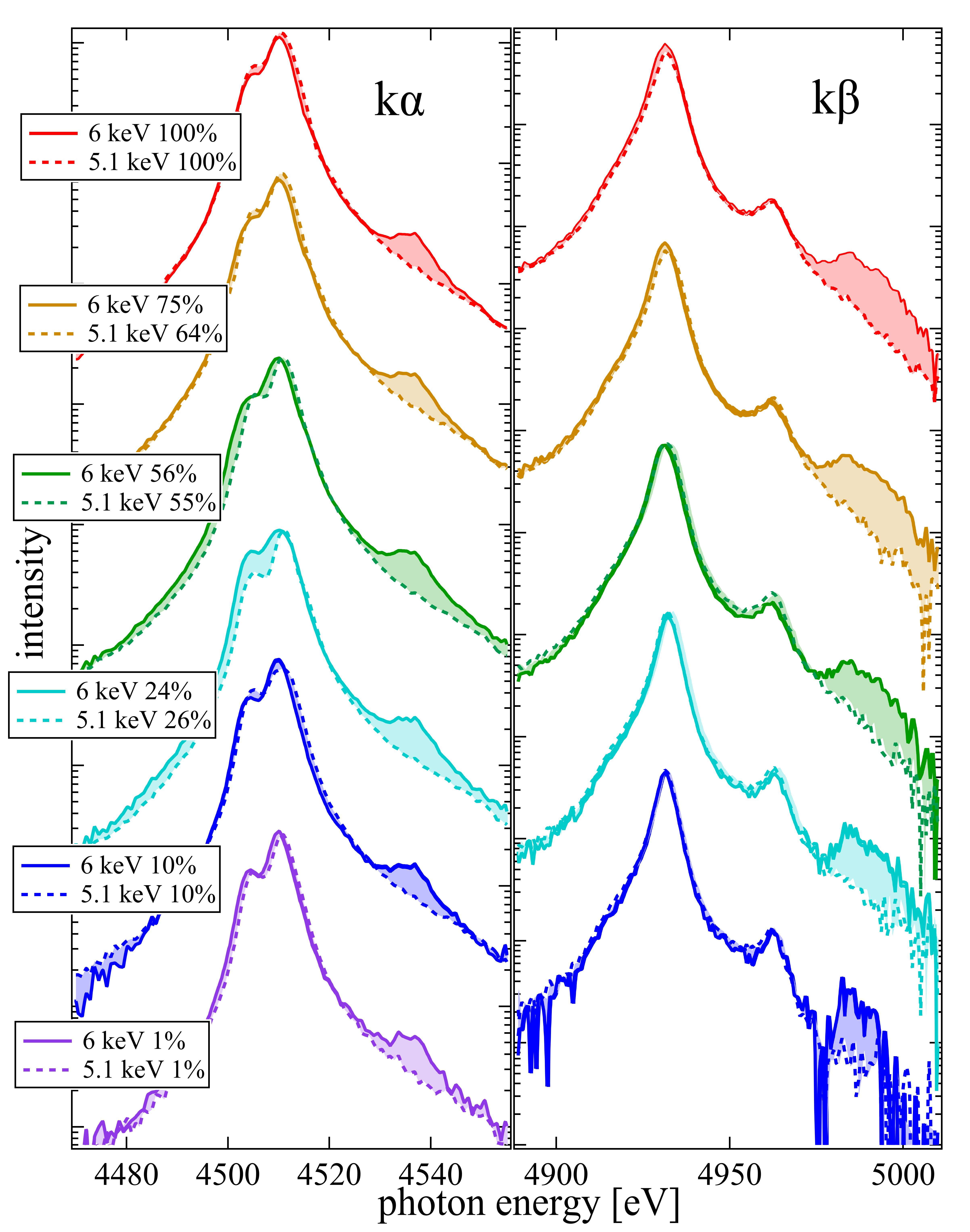} 
\caption{XFEL induced fluorescent spectra with a range of attenuators for both K$\alpha$ and K$\beta$ emission for XFEL photon energies of  5.1 keV and 6 keV. Spectra have been offset vertically for comparison. The shaded areas highlight fill the difference between the 6 and 5.1 keV XFEL photon energy. The L-shell satellite intensity is clearly visible over a wide intensity range in both the K$\alpha$ and the K$\beta$ peak for the 6 keV photon energy, but not for the 5.1 keV photon energy.  \label{fig:intensity_comp_ka_kb}.}
\end{figure}

The recorded x-ray spectra are a result of fluorescent spectrum of the K-shell hole that is created with XFEL heating (as both photon energies used are above the K-edge). Fluorescent recombination of the $2p-1s$ transition results in the K$\alpha$ peak, and the $3p-1s$ and continuum to $1s$ transition results in the K$\beta$ spectrum. Spectra for a range of attenuations, giving peak intensities spanning two orders of magnitude are presented in Fig. \ref{fig:intensity_comp_ka_kb}. Two main observations are apparent: the L-shell satellite only appears for 6 keV, and its magnitude is largely independent of the intensity of the FEL and therefore the plasma temperature within the lifetime of the emission.

The observation of L-shell vacancies in the emission spectra for the higher photon energy of 6 keV is indicative of shake-off process as the energy for both K and L-shell ionisation must be carried by the incoming XFEL photon, whereas absent for the 5.1 keV energy as the photon does not have enough energy for K and L shell ionisation. However, vacancies in the L-shell can be created through other mechanisms, namely: Auger emission, photoionisation, collisional ionisation, excitation and shake processes, all of which occur in XFEL heated plasmas to varying degrees as a function of time. To model the emission spectra, we use the collisional radiative code \texttt{BigBart}, that calculates the ionisation and recombination processes as a function time while accounting for the non-thermal electron distribution and degenerate electron populations \cite{Varga2013}. To confirm our hypothesis as to whether shake processes persist into the dense plasma regime, we calculate the emission spectra both including and excluding shake processes to identify the dominant contribution to creating L-shell vacancies. Shake processes have been included in the BigBart code through the approach of Carlson and Nestor, which calculates the probability from a tractable first principles approach that matches well with data \cite{Carlson1973}. The cut-off between shake-up (the excitation equivalent to shake-off) and shake-off is defined as the beginning of the continuum.

\begin{figure}
\includegraphics[scale=0.42]{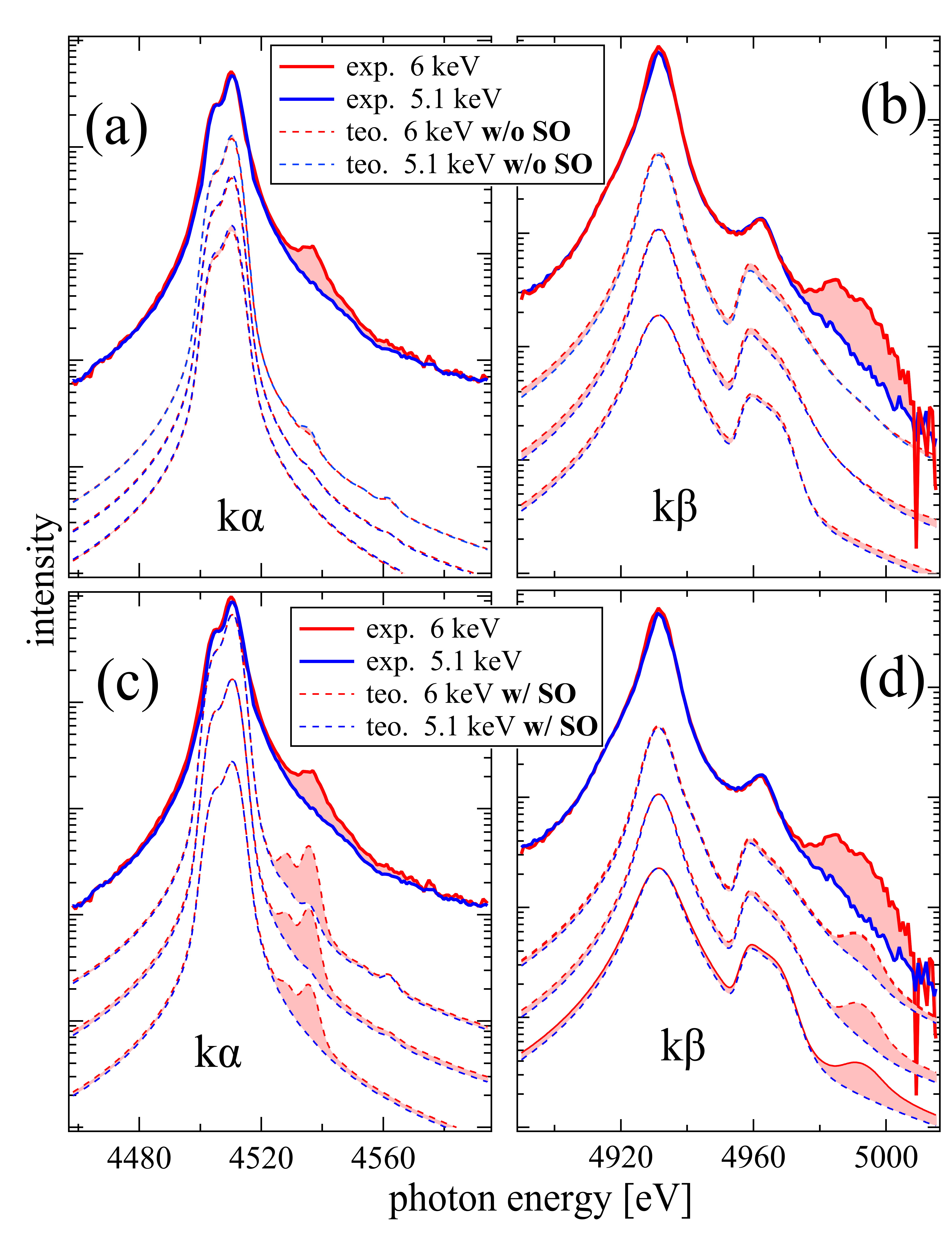} 
\caption{Experimental data obtained at full intensity for XFEL photon energies of 5.1 keV and 6 keV compared to models without shake-off (w/o SO) (a and b), and with shake-off (w/ SO) (c and d). The calculations were done for three XFEL intensities (0.1, 0.5 and 1$\times10^{16}$ W cm$^{-2}$) increasing in intensity vertically. The shaded areas highlight fill the difference between the 6 and 5.1 keV XFEL photon energy. \label{fig:Exp_teo_w_wo_shake}.}
\end{figure}

In Fig. \ref{fig:Exp_teo_w_wo_shake}, we compare the experimental data at the highest intensity with calculations for three intensities using the established XFEL fluorescence modelling, i.e. without shake processes, (Fig. \ref{fig:Exp_teo_w_wo_shake}a and b), and with shake processes included (Fig. \ref{fig:Exp_teo_w_wo_shake}c and d). For the calculations without shake in Fig. \ref{fig:Exp_teo_w_wo_shake}(a and b), we see a relatively minor L-shell satellite in the K$\alpha$ and the K$\beta$ spectra compared to the experiment. There is a slight increase in the L-shell satellite in the K$\alpha$ spectra for higher intensities in (Fig. \ref{fig:Exp_teo_w_wo_shake}a), which is due to increased collisional ionisation due to higher temperatures. However, the magnitude of the difference between 5.1 keV and 6 keV for the calculated spectra for both the K$\alpha$ and the K$\beta$ is minor when compared to experiment in Fig. \ref{fig:Exp_teo_w_wo_shake}a and b. This is in stark contrast to the calculations including shake processes in Fig. \ref{fig:Exp_teo_w_wo_shake}c and d, where the L-shell satellite is much more prominent for the 6 keV than the 5.1 keV photon energies for both the K$\alpha$ and the K$\beta$ spectra. The calculated spectra are in close agreement with our experimental results that show a minimal dependance on intensity, whereas a distinct sensitivity to photon energy. This is consistent with a shake-off ionisation process, wherein a rapid double ionisation of the K and L-shell electrons occurs, and subsequent fluorescent recombination to the K-shell vacancy results in an L-shell satellite. Other studies have measured K$\alpha^1$ (first L vacancy satellite) to K$\alpha^0$ (cold  K$\alpha$) ratios of a few percent for shake-off in cold titanium \cite{Murti2008,Oura2002} and calculations predict similar probabilities \cite{Mukoyama1987}. To gauge the impact of uncertainties in the collisional ionisation rates, we have altered the M and L-shell cross-sections and found only slight changes in the satellite peaks for both photon energies and were unable to replicate the experimental results. The simulations also show that shake-off is by far the dominant process when compared to shake-up due o the lack of final states available for excitation in these conditions.

In conclusion, we have presented experimental evidence, supported by calculations, of shake-off occurring in dense plasmas. Shake-off has been shown to contribute to prominent  K$\alpha$ satellites in solid matter elsewhere, and we have shown that this process persists into the plasma regime unperturbed for electron temperatures up to 10 eV. This work shows unambiguously that including shake processes in atomic and collisional radiative codes is essential for experiments that use satellites as a fingerprint of temperature, density or ion charge state distributions. Furthermore, shake processes occur more readily for higher shells creating holes and hence satellites in the K$\alpha$ and the K$\beta$ spectra (although beyond the experimental resolution of this study). Shake process in higher shells has generally a higher probability and is less demanding on the initial photon energy (less energy needed for ionisation), further highlighting the need to include shake processes in the interpretation of emission spectra in plasma experiments. 

The results also show the initial ejected electron survives unperturbed in the plasma for the duration of the shake processes. With timescales on the order of 10’s to 100’s of attoseconds (as), shake-off processes are much shorter than the timescale of plasma oscillations or elastic electron collisions in the continuum and could be used as a diagnostic for ionisation potentials in varying densities and temperatures. Furthermore, XFELs with  pulse durations of a few femtoseconds could allow us to distinguish between satellites caused by shake-off (as) and those created by collisional ionisation.

\begin{acknowledgments}
We acknowledge the European XFEL in Schenefeld, Germany, for provision of X-ray free-electron laser beamtime at the Scientific Instrument HED (High Energy Density Science) under Proposal No. 2586 and would like to thank the staff for their assistance. The authors are grateful to the HIBEF user consortium for the provision of instrumentation and staff that enabled this experiment. Data recorded for the experiment at the European XFEL are available at doi:10.22003/XFEL.EU-DATA-002586-00. This work was supported by FCT Concurso de Projetos de I\&D em Todos os Dom\'{i}nios Cient\'{i}ficos: 2022.09213.PTDC. FCT Concurso Est\'{i}mulo ao Emprego Cient\'{i}fico Individual CEECIND (DOI:10.54499/2022.00804.CEECIND/CP1713/CT0003), and the for Call Advanced Computer Projects: 2024.07903.CPCA.A2. This work received support from then NanoXCan project, a European Council Pathfinder Project 101047223. L. A. acknowledges support from FCT (Foundation for Science and Technology - Portugal) under Grant No. UI/BD/153734/2022. P.E  acknowledges support from FCT under grant no. PD/BD/150410/2019. P. V. and L. A. acknowledge the support of the Spanish Government through the project PID2021-124129OB-I00 funded by MCIN/AEI/10.13039/501100011033/ERDF. The authors gratefully acknowledge the Universidad Polit\'{e}cnica de Madrid (www.upm.es) for providing computing resources on Magerit Supercomputer.

\end{acknowledgments}

%


\begin{thebibliography}{0}%
\makeatletter
\providecommand \@ifxundefined [1]{%
 \@ifx{#1\undefined}
}%
\providecommand \@ifnum [1]{%
 \ifnum #1\expandafter \@firstoftwo
 \else \expandafter \@secondoftwo
 \fi
}%
\providecommand \@ifx [1]{%
 \ifx #1\expandafter \@firstoftwo
 \else \expandafter \@secondoftwo
 \fi
}%
\providecommand \natexlab [1]{#1}%
\providecommand \enquote  [1]{``#1''}%
\providecommand \bibnamefont  [1]{#1}%
\providecommand \bibfnamefont [1]{#1}%
\providecommand \citenamefont [1]{#1}%
\providecommand \href@noop [0]{\@secondoftwo}%
\providecommand \href [0]{\begingroup \@sanitize@url \@href}%
\providecommand \@href[1]{\@@startlink{#1}\@@href}%
\providecommand \@@href[1]{\endgroup#1\@@endlink}%
\providecommand \@sanitize@url [0]{\catcode `\\12\catcode `\$12\catcode
  `\&12\catcode `\#12\catcode `\^12\catcode `\_12\catcode `\%12\relax}%
\providecommand \@@startlink[1]{}%
\providecommand \@@endlink[0]{}%
\providecommand \url  [0]{\begingroup\@sanitize@url \@url }%
\providecommand \@url [1]{\endgroup\@href {#1}{\urlprefix }}%
\providecommand \urlprefix  [0]{URL }%
\providecommand \Eprint [0]{\href }%
\providecommand \doibase [0]{https://doi.org/}%
\providecommand \selectlanguage [0]{\@gobble}%
\providecommand \bibinfo  [0]{\@secondoftwo}%
\providecommand \bibfield  [0]{\@secondoftwo}%
\providecommand \translation [1]{[#1]}%
\providecommand \BibitemOpen [0]{}%
\providecommand \bibitemStop [0]{}%
\providecommand \bibitemNoStop [0]{.\EOS\space}%
\providecommand \EOS [0]{\spacefactor3000\relax}%
\providecommand \BibitemShut  [1]{\csname bibitem#1\endcsname}%
\let\auto@bib@innerbib\@empty
\end{thebibliography}%


\begin{thebibliography}{20}%
\makeatletter
\providecommand \@ifxundefined [1]{%
 \@ifx{#1\undefined}
}%
\providecommand \@ifnum [1]{%
 \ifnum #1\expandafter \@firstoftwo
 \else \expandafter \@secondoftwo
 \fi
}%
\providecommand \@ifx [1]{%
 \ifx #1\expandafter \@firstoftwo
 \else \expandafter \@secondoftwo
 \fi
}%
\providecommand \natexlab [1]{#1}%
\providecommand \enquote  [1]{``#1''}%
\providecommand \bibnamefont  [1]{#1}%
\providecommand \bibfnamefont [1]{#1}%
\providecommand \citenamefont [1]{#1}%
\providecommand \href@noop [0]{\@secondoftwo}%
\providecommand \href [0]{\begingroup \@sanitize@url \@href}%
\providecommand \@href[1]{\@@startlink{#1}\@@href}%
\providecommand \@@href[1]{\endgroup#1\@@endlink}%
\providecommand \@sanitize@url [0]{\catcode `\\12\catcode `\$12\catcode
  `\&12\catcode `\#12\catcode `\^12\catcode `\_12\catcode `\%12\relax}%
\providecommand \@@startlink[1]{}%
\providecommand \@@endlink[0]{}%
\providecommand \url  [0]{\begingroup\@sanitize@url \@url }%
\providecommand \@url [1]{\endgroup\@href {#1}{\urlprefix }}%
\providecommand \urlprefix  [0]{URL }%
\providecommand \Eprint [0]{\href }%
\providecommand \doibase [0]{https://doi.org/}%
\providecommand \selectlanguage [0]{\@gobble}%
\providecommand \bibinfo  [0]{\@secondoftwo}%
\providecommand \bibfield  [0]{\@secondoftwo}%
\providecommand \translation [1]{[#1]}%
\providecommand \BibitemOpen [0]{}%
\providecommand \bibitemStop [0]{}%
\providecommand \bibitemNoStop [0]{.\EOS\space}%
\providecommand \EOS [0]{\spacefactor3000\relax}%
\providecommand \BibitemShut  [1]{\csname bibitem#1\endcsname}%
\let\auto@bib@innerbib\@empty
\bibitem [{\citenamefont {Vinko}\ \emph {et~al.}(2012)\citenamefont {Vinko},
  \citenamefont {Ciricosta}, \citenamefont {Cho}, \citenamefont {Engelhorn},
  \citenamefont {Chung}, \citenamefont {Brown}, \citenamefont {Burian},
  \citenamefont {Chalupsk{\`y}}, \citenamefont {Falcone}, \citenamefont
  {Graves} \emph {et~al.}}]{Vinko2012}%
  \BibitemOpen
  \bibfield  {author} {\bibinfo {author} {\bibfnamefont {S.}~\bibnamefont
  {Vinko}}, \bibinfo {author} {\bibfnamefont {O.}~\bibnamefont {Ciricosta}},
  \bibinfo {author} {\bibfnamefont {B.}~\bibnamefont {Cho}}, \bibinfo {author}
  {\bibfnamefont {K.}~\bibnamefont {Engelhorn}}, \bibinfo {author}
  {\bibfnamefont {H.-K.}\ \bibnamefont {Chung}}, \bibinfo {author}
  {\bibfnamefont {C.}~\bibnamefont {Brown}}, \bibinfo {author} {\bibfnamefont
  {T.}~\bibnamefont {Burian}}, \bibinfo {author} {\bibfnamefont
  {J.}~\bibnamefont {Chalupsk{\`y}}}, \bibinfo {author} {\bibfnamefont
  {R.}~\bibnamefont {Falcone}}, \bibinfo {author} {\bibfnamefont
  {C.}~\bibnamefont {Graves}}, \emph {et~al.},\ }\href@noop {} {\ \textbf
  {\bibinfo {volume} {482}},\ \bibinfo {pages} {59} (\bibinfo {year}
  {2012})}\BibitemShut {NoStop}%
\bibitem [{\citenamefont {Fletcher}\ \emph {et~al.}(2015)\citenamefont
  {Fletcher}, \citenamefont {Lee}, \citenamefont {D{\"o}ppner}, \citenamefont
  {Galtier}, \citenamefont {Nagler}, \citenamefont {Heimann}, \citenamefont
  {Fortmann}, \citenamefont {LePape}, \citenamefont {Ma}, \citenamefont
  {Millot} \emph {et~al.}}]{Fletcher2015}%
  \BibitemOpen
  \bibfield  {author} {\bibinfo {author} {\bibfnamefont {L.}~\bibnamefont
  {Fletcher}}, \bibinfo {author} {\bibfnamefont {H.}~\bibnamefont {Lee}},
  \bibinfo {author} {\bibfnamefont {T.}~\bibnamefont {D{\"o}ppner}}, \bibinfo
  {author} {\bibfnamefont {E.}~\bibnamefont {Galtier}}, \bibinfo {author}
  {\bibfnamefont {B.}~\bibnamefont {Nagler}}, \bibinfo {author} {\bibfnamefont
  {P.}~\bibnamefont {Heimann}}, \bibinfo {author} {\bibfnamefont
  {C.}~\bibnamefont {Fortmann}}, \bibinfo {author} {\bibfnamefont
  {S.}~\bibnamefont {LePape}}, \bibinfo {author} {\bibfnamefont
  {T.}~\bibnamefont {Ma}}, \bibinfo {author} {\bibfnamefont {M.}~\bibnamefont
  {Millot}}, \emph {et~al.},\ }\href@noop {} {\ \textbf {\bibinfo {volume}
  {9}},\ \bibinfo {pages} {274} (\bibinfo {year} {2015})}\BibitemShut {NoStop}%
\bibitem [{\citenamefont {Laso~Garcia}\ \emph {et~al.}(2024)\citenamefont
  {Laso~Garcia}, \citenamefont {Yang}, \citenamefont {Bouffetier},
  \citenamefont {Appel}, \citenamefont {Baehtz}, \citenamefont {Hagemann},
  \citenamefont {H{\"o}ppner}, \citenamefont {Humphries}, \citenamefont
  {Kluge}, \citenamefont {Mishchenko} \emph {et~al.}}]{LasoGarcia2024}%
  \BibitemOpen
  \bibfield  {author} {\bibinfo {author} {\bibfnamefont {A.}~\bibnamefont
  {Laso~Garcia}}, \bibinfo {author} {\bibfnamefont {L.}~\bibnamefont {Yang}},
  \bibinfo {author} {\bibfnamefont {V.}~\bibnamefont {Bouffetier}}, \bibinfo
  {author} {\bibfnamefont {K.}~\bibnamefont {Appel}}, \bibinfo {author}
  {\bibfnamefont {C.}~\bibnamefont {Baehtz}}, \bibinfo {author} {\bibfnamefont
  {J.}~\bibnamefont {Hagemann}}, \bibinfo {author} {\bibfnamefont
  {H.}~\bibnamefont {H{\"o}ppner}}, \bibinfo {author} {\bibfnamefont
  {O.}~\bibnamefont {Humphries}}, \bibinfo {author} {\bibfnamefont
  {T.}~\bibnamefont {Kluge}}, \bibinfo {author} {\bibfnamefont
  {M.}~\bibnamefont {Mishchenko}}, \emph {et~al.},\ }\href@noop {} {\bibfield
  {journal} {\bibinfo  {journal} {Nature Communications}\ }\textbf {\bibinfo
  {volume} {15}},\ \bibinfo {pages} {7896} (\bibinfo {year}
  {2024})}\BibitemShut {NoStop}%
\bibitem [{\citenamefont {Levy}\ \emph {et~al.}(2015)\citenamefont {Levy},
  \citenamefont {Audebert}, \citenamefont {Shepherd}, \citenamefont {Dunn},
  \citenamefont {Cammarata}, \citenamefont {Ciricosta}, \citenamefont
  {Deneuville}, \citenamefont {Dorchies}, \citenamefont {Fajardo},
  \citenamefont {Fourment} \emph {et~al.}}]{Levy2015}%
  \BibitemOpen
  \bibfield  {author} {\bibinfo {author} {\bibfnamefont {A.}~\bibnamefont
  {Levy}}, \bibinfo {author} {\bibfnamefont {P.}~\bibnamefont {Audebert}},
  \bibinfo {author} {\bibfnamefont {R.}~\bibnamefont {Shepherd}}, \bibinfo
  {author} {\bibfnamefont {J.}~\bibnamefont {Dunn}}, \bibinfo {author}
  {\bibfnamefont {M.}~\bibnamefont {Cammarata}}, \bibinfo {author}
  {\bibfnamefont {O.}~\bibnamefont {Ciricosta}}, \bibinfo {author}
  {\bibfnamefont {F.}~\bibnamefont {Deneuville}}, \bibinfo {author}
  {\bibfnamefont {F.}~\bibnamefont {Dorchies}}, \bibinfo {author}
  {\bibfnamefont {M.}~\bibnamefont {Fajardo}}, \bibinfo {author} {\bibfnamefont
  {C.}~\bibnamefont {Fourment}}, \emph {et~al.},\ }\href@noop {} {\ \textbf
  {\bibinfo {volume} {22}},\ \bibinfo {pages} {030703} (\bibinfo {year}
  {2015})}\BibitemShut {NoStop}%
\bibitem [{\citenamefont {Ciricosta}\ \emph {et~al.}(2012)\citenamefont
  {Ciricosta}, \citenamefont {Vinko}, \citenamefont {Chung}, \citenamefont
  {Cho}, \citenamefont {Brown}, \citenamefont {Burian}, \citenamefont
  {Chalupsk\'y}, \citenamefont {Engelhorn}, \citenamefont {Falcone},
  \citenamefont {Graves}, \citenamefont {H\'ajkov\'a}, \citenamefont
  {Higginbotham}, \citenamefont {Juha}, \citenamefont {Krzywinski},
  \citenamefont {Lee}, \citenamefont {Messerschmidt}, \citenamefont {Murphy},
  \citenamefont {Ping}, \citenamefont {Rackstraw}, \citenamefont {Scherz},
  \citenamefont {Schlotter}, \citenamefont {Toleikis}, \citenamefont {Turner},
  \citenamefont {Vysin}, \citenamefont {Wang}, \citenamefont {Wu},
  \citenamefont {Zastrau}, \citenamefont {Zhu}, \citenamefont {Lee},
  \citenamefont {Heimann}, \citenamefont {Nagler},\ and\ \citenamefont
  {Wark}}]{Ciricosta2012}%
  \BibitemOpen
  \bibfield  {author} {\bibinfo {author} {\bibfnamefont {O.}~\bibnamefont
  {Ciricosta}}, \bibinfo {author} {\bibfnamefont {S.~M.}\ \bibnamefont
  {Vinko}}, \bibinfo {author} {\bibfnamefont {H.-K.}\ \bibnamefont {Chung}},
  \bibinfo {author} {\bibfnamefont {B.~I.}\ \bibnamefont {Cho}}, \bibinfo
  {author} {\bibfnamefont {C.~R.~D.}\ \bibnamefont {Brown}}, \bibinfo {author}
  {\bibfnamefont {T.}~\bibnamefont {Burian}}, \bibinfo {author} {\bibfnamefont
  {J.}~\bibnamefont {Chalupsk\'y}}, \bibinfo {author} {\bibfnamefont
  {K.}~\bibnamefont {Engelhorn}}, \bibinfo {author} {\bibfnamefont {R.~W.}\
  \bibnamefont {Falcone}}, \bibinfo {author} {\bibfnamefont {C.}~\bibnamefont
  {Graves}}, \bibinfo {author} {\bibfnamefont {V.}~\bibnamefont {H\'ajkov\'a}},
  \bibinfo {author} {\bibfnamefont {A.}~\bibnamefont {Higginbotham}}, \bibinfo
  {author} {\bibfnamefont {L.}~\bibnamefont {Juha}}, \bibinfo {author}
  {\bibfnamefont {J.}~\bibnamefont {Krzywinski}}, \bibinfo {author}
  {\bibfnamefont {H.~J.}\ \bibnamefont {Lee}}, \bibinfo {author} {\bibfnamefont
  {M.}~\bibnamefont {Messerschmidt}}, \bibinfo {author} {\bibfnamefont {C.~D.}\
  \bibnamefont {Murphy}}, \bibinfo {author} {\bibfnamefont {Y.}~\bibnamefont
  {Ping}}, \bibinfo {author} {\bibfnamefont {D.~S.}\ \bibnamefont {Rackstraw}},
  \bibinfo {author} {\bibfnamefont {A.}~\bibnamefont {Scherz}}, \bibinfo
  {author} {\bibfnamefont {W.}~\bibnamefont {Schlotter}}, \bibinfo {author}
  {\bibfnamefont {S.}~\bibnamefont {Toleikis}}, \bibinfo {author}
  {\bibfnamefont {J.~J.}\ \bibnamefont {Turner}}, \bibinfo {author}
  {\bibfnamefont {L.}~\bibnamefont {Vysin}}, \bibinfo {author} {\bibfnamefont
  {T.}~\bibnamefont {Wang}}, \bibinfo {author} {\bibfnamefont {B.}~\bibnamefont
  {Wu}}, \bibinfo {author} {\bibfnamefont {U.}~\bibnamefont {Zastrau}},
  \bibinfo {author} {\bibfnamefont {D.}~\bibnamefont {Zhu}}, \bibinfo {author}
  {\bibfnamefont {R.~W.}\ \bibnamefont {Lee}}, \bibinfo {author} {\bibfnamefont
  {P.}~\bibnamefont {Heimann}}, \bibinfo {author} {\bibfnamefont
  {B.}~\bibnamefont {Nagler}},\ and\ \bibinfo {author} {\bibfnamefont {J.~S.}\
  \bibnamefont {Wark}},\ }\href
  {https://doi.org/10.1103/PhysRevLett.109.065002} {\ \textbf {\bibinfo
  {volume} {109}},\ \bibinfo {pages} {065002} (\bibinfo {year}
  {2012})}\BibitemShut {NoStop}%
\bibitem [{\citenamefont {van~den Berg}\ \emph {et~al.}(2018)\citenamefont
  {van~den Berg}, \citenamefont {Fernandez-Tello}, \citenamefont {Burian},
  \citenamefont {Chalupsk\'y}, \citenamefont {Chung}, \citenamefont
  {Ciricosta}, \citenamefont {Dakovski}, \citenamefont {H\'ajkov\'a},
  \citenamefont {Hollebon}, \citenamefont {Juha}, \citenamefont {Krzywinski},
  \citenamefont {Lee}, \citenamefont {Minitti}, \citenamefont {Preston},
  \citenamefont {de~la Varga}, \citenamefont {Vozda}, \citenamefont {Zastrau},
  \citenamefont {Wark}, \citenamefont {Velarde},\ and\ \citenamefont
  {Vinko}}]{Berg2018}%
  \BibitemOpen
  \bibfield  {author} {\bibinfo {author} {\bibfnamefont {Q.~Y.}\ \bibnamefont
  {van~den Berg}}, \bibinfo {author} {\bibfnamefont {E.~V.}\ \bibnamefont
  {Fernandez-Tello}}, \bibinfo {author} {\bibfnamefont {T.}~\bibnamefont
  {Burian}}, \bibinfo {author} {\bibfnamefont {J.}~\bibnamefont {Chalupsk\'y}},
  \bibinfo {author} {\bibfnamefont {H.-K.}\ \bibnamefont {Chung}}, \bibinfo
  {author} {\bibfnamefont {O.}~\bibnamefont {Ciricosta}}, \bibinfo {author}
  {\bibfnamefont {G.~L.}\ \bibnamefont {Dakovski}}, \bibinfo {author}
  {\bibfnamefont {V.}~\bibnamefont {H\'ajkov\'a}}, \bibinfo {author}
  {\bibfnamefont {P.}~\bibnamefont {Hollebon}}, \bibinfo {author}
  {\bibfnamefont {L.}~\bibnamefont {Juha}}, \bibinfo {author} {\bibfnamefont
  {J.}~\bibnamefont {Krzywinski}}, \bibinfo {author} {\bibfnamefont {R.~W.}\
  \bibnamefont {Lee}}, \bibinfo {author} {\bibfnamefont {M.~P.}\ \bibnamefont
  {Minitti}}, \bibinfo {author} {\bibfnamefont {T.~R.}\ \bibnamefont
  {Preston}}, \bibinfo {author} {\bibfnamefont {A.~G.}\ \bibnamefont {de~la
  Varga}}, \bibinfo {author} {\bibfnamefont {V.}~\bibnamefont {Vozda}},
  \bibinfo {author} {\bibfnamefont {U.}~\bibnamefont {Zastrau}}, \bibinfo
  {author} {\bibfnamefont {J.~S.}\ \bibnamefont {Wark}}, \bibinfo {author}
  {\bibfnamefont {P.}~\bibnamefont {Velarde}},\ and\ \bibinfo {author}
  {\bibfnamefont {S.~M.}\ \bibnamefont {Vinko}},\ }\href
  {https://doi.org/10.1103/PhysRevLett.120.055002} {\ \textbf {\bibinfo
  {volume} {120}},\ \bibinfo {pages} {055002} (\bibinfo {year}
  {2018})}\BibitemShut {NoStop}%
\bibitem [{\citenamefont {Snell}\ and\ \citenamefont
  {Pleasonton}(1957)}]{Snell1957}%
  \BibitemOpen
  \bibfield  {author} {\bibinfo {author} {\bibfnamefont {A.~H.}\ \bibnamefont
  {Snell}}\ and\ \bibinfo {author} {\bibfnamefont {F.}~\bibnamefont
  {Pleasonton}},\ }\href@noop {} {\bibfield  {journal} {\bibinfo  {journal}
  {Physical Review}\ }\textbf {\bibinfo {volume} {107}},\ \bibinfo {pages}
  {740} (\bibinfo {year} {1957})}\BibitemShut {NoStop}%
\bibitem [{\citenamefont {Carlson}\ and\ \citenamefont
  {Krause}(1965)}]{Carlson1965}%
  \BibitemOpen
  \bibfield  {author} {\bibinfo {author} {\bibfnamefont {T.~A.}\ \bibnamefont
  {Carlson}}\ and\ \bibinfo {author} {\bibfnamefont {M.~O.}\ \bibnamefont
  {Krause}},\ }\href {https://doi.org/10.1103/PhysRev.140.A1057} {\bibfield
  {journal} {\bibinfo  {journal} {Phys. Rev.}\ }\textbf {\bibinfo {volume}
  {140}},\ \bibinfo {pages} {A1057} (\bibinfo {year} {1965})}\BibitemShut
  {NoStop}%
\bibitem [{\citenamefont {Thomas}(1984)}]{Thomas1984}%
  \BibitemOpen
  \bibfield  {author} {\bibinfo {author} {\bibfnamefont {T.~D.}\ \bibnamefont
  {Thomas}},\ }\href {https://doi.org/10.1103/PhysRevLett.52.417} {\bibfield
  {journal} {\bibinfo  {journal} {Phys. Rev. Lett.}\ }\textbf {\bibinfo
  {volume} {52}},\ \bibinfo {pages} {417} (\bibinfo {year} {1984})}\BibitemShut
  {NoStop}%
\bibitem [{\citenamefont {Carlson}\ and\ \citenamefont
  {Nestor}(1973)}]{Carlson1973}%
  \BibitemOpen
  \bibfield  {author} {\bibinfo {author} {\bibfnamefont {T.~A.}\ \bibnamefont
  {Carlson}}\ and\ \bibinfo {author} {\bibfnamefont {C.~W.}\ \bibnamefont
  {Nestor}},\ }\href {https://doi.org/10.1103/PhysRevA.8.2887} {\bibfield
  {journal} {\bibinfo  {journal} {Phys. Rev. A}\ }\textbf {\bibinfo {volume}
  {8}},\ \bibinfo {pages} {2887} (\bibinfo {year} {1973})}\BibitemShut
  {NoStop}%
\bibitem [{\citenamefont {\AA{}berg}\ \emph {et~al.}(1976)\citenamefont
  {\AA{}berg}, \citenamefont {Jamison},\ and\ \citenamefont
  {Richard}}]{Aaberg1976}%
  \BibitemOpen
  \bibfield  {author} {\bibinfo {author} {\bibfnamefont {T.}~\bibnamefont
  {\AA{}berg}}, \bibinfo {author} {\bibfnamefont {K.~A.}\ \bibnamefont
  {Jamison}},\ and\ \bibinfo {author} {\bibfnamefont {P.}~\bibnamefont
  {Richard}},\ }\href {https://doi.org/10.1103/PhysRevLett.37.63} {\bibfield
  {journal} {\bibinfo  {journal} {Phys. Rev. Lett.}\ }\textbf {\bibinfo
  {volume} {37}},\ \bibinfo {pages} {63} (\bibinfo {year} {1976})}\BibitemShut
  {NoStop}%
\bibitem [{\citenamefont {Mukoyama}\ and\ \citenamefont
  {Taniguchi}(1987)}]{Mukoyama1987}%
  \BibitemOpen
  \bibfield  {author} {\bibinfo {author} {\bibfnamefont {T.}~\bibnamefont
  {Mukoyama}}\ and\ \bibinfo {author} {\bibfnamefont {K.}~\bibnamefont
  {Taniguchi}},\ }\href {https://doi.org/10.1103/PhysRevA.36.693} {\bibfield
  {journal} {\bibinfo  {journal} {Phys. Rev. A}\ }\textbf {\bibinfo {volume}
  {36}},\ \bibinfo {pages} {693} (\bibinfo {year} {1987})}\BibitemShut
  {NoStop}%
\bibitem [{\citenamefont {Oura}\ \emph {et~al.}(2003)\citenamefont {Oura},
  \citenamefont {Mukoyama}, \citenamefont {Taguchi}, \citenamefont {Takeuchi},
  \citenamefont {Haruna},\ and\ \citenamefont {Shin}}]{Oura2003}%
  \BibitemOpen
  \bibfield  {author} {\bibinfo {author} {\bibfnamefont {M.}~\bibnamefont
  {Oura}}, \bibinfo {author} {\bibfnamefont {T.}~\bibnamefont {Mukoyama}},
  \bibinfo {author} {\bibfnamefont {M.}~\bibnamefont {Taguchi}}, \bibinfo
  {author} {\bibfnamefont {T.}~\bibnamefont {Takeuchi}}, \bibinfo {author}
  {\bibfnamefont {T.}~\bibnamefont {Haruna}},\ and\ \bibinfo {author}
  {\bibfnamefont {S.}~\bibnamefont {Shin}},\ }\href@noop {} {\bibfield
  {journal} {\bibinfo  {journal} {Physical review letters}\ }\textbf {\bibinfo
  {volume} {90}},\ \bibinfo {pages} {173002} (\bibinfo {year}
  {2003})}\BibitemShut {NoStop}%
\bibitem [{\citenamefont {Zastrau}\ \emph {et~al.}(2021)\citenamefont
  {Zastrau}, \citenamefont {Appel}, \citenamefont {Baehtz}, \citenamefont
  {Baehr}, \citenamefont {Batchelor}, \citenamefont {Bergh{\"a}user},
  \citenamefont {Banjafar}, \citenamefont {Brambrink}, \citenamefont
  {Cerantola}, \citenamefont {Cowan} \emph {et~al.}}]{Zastrau2021}%
  \BibitemOpen
  \bibfield  {author} {\bibinfo {author} {\bibfnamefont {U.}~\bibnamefont
  {Zastrau}}, \bibinfo {author} {\bibfnamefont {K.}~\bibnamefont {Appel}},
  \bibinfo {author} {\bibfnamefont {C.}~\bibnamefont {Baehtz}}, \bibinfo
  {author} {\bibfnamefont {O.}~\bibnamefont {Baehr}}, \bibinfo {author}
  {\bibfnamefont {L.}~\bibnamefont {Batchelor}}, \bibinfo {author}
  {\bibfnamefont {A.}~\bibnamefont {Bergh{\"a}user}}, \bibinfo {author}
  {\bibfnamefont {M.}~\bibnamefont {Banjafar}}, \bibinfo {author}
  {\bibfnamefont {E.}~\bibnamefont {Brambrink}}, \bibinfo {author}
  {\bibfnamefont {V.}~\bibnamefont {Cerantola}}, \bibinfo {author}
  {\bibfnamefont {T.~E.}\ \bibnamefont {Cowan}}, \emph {et~al.},\ }\href@noop
  {} {\bibfield  {journal} {\bibinfo  {journal} {Journal of synchrotron
  radiation}\ }\textbf {\bibinfo {volume} {28}},\ \bibinfo {pages} {1393}
  (\bibinfo {year} {2021})}\BibitemShut {NoStop}%
\bibitem [{\citenamefont {Preston}\ \emph {et~al.}(2020)\citenamefont
  {Preston}, \citenamefont {G{\"o}de}, \citenamefont {Schwinkendorf},
  \citenamefont {Appel}, \citenamefont {Brambrink}, \citenamefont {Cerantola},
  \citenamefont {H{\"o}ppner}, \citenamefont {Makita}, \citenamefont {Pelka},
  \citenamefont {Prescher} \emph {et~al.}}]{Preston2020}%
  \BibitemOpen
  \bibfield  {author} {\bibinfo {author} {\bibfnamefont {T.}~\bibnamefont
  {Preston}}, \bibinfo {author} {\bibfnamefont {S.}~\bibnamefont {G{\"o}de}},
  \bibinfo {author} {\bibfnamefont {J.-P.}\ \bibnamefont {Schwinkendorf}},
  \bibinfo {author} {\bibfnamefont {K.}~\bibnamefont {Appel}}, \bibinfo
  {author} {\bibfnamefont {E.}~\bibnamefont {Brambrink}}, \bibinfo {author}
  {\bibfnamefont {V.}~\bibnamefont {Cerantola}}, \bibinfo {author}
  {\bibfnamefont {H.}~\bibnamefont {H{\"o}ppner}}, \bibinfo {author}
  {\bibfnamefont {M.}~\bibnamefont {Makita}}, \bibinfo {author} {\bibfnamefont
  {A.}~\bibnamefont {Pelka}}, \bibinfo {author} {\bibfnamefont
  {C.}~\bibnamefont {Prescher}}, \emph {et~al.},\ }\href
  {https://doi.org/10.1088/1748-0221/15/11/P11033} {\bibfield  {journal}
  {\bibinfo  {journal} {Journal of Instrumentation}\ }\textbf {\bibinfo
  {volume} {15}}\bibinfo  {number} { (11)},\ \bibinfo {pages}
  {P11033}}\BibitemShut {NoStop}%
\bibitem [{\citenamefont {Chalupsk\'{y}}\ \emph {et~al.}(2013)\citenamefont
  {Chalupsk\'{y}}, \citenamefont {Burian}, \citenamefont {H\'{a}jkov\'{a}},
  \citenamefont {Juha}, \citenamefont {Polcar}, \citenamefont {Gaudin},
  \citenamefont {Nagasono}, \citenamefont {Sobierajski}, \citenamefont
  {Yabashi},\ and\ \citenamefont {Krzywinski}}]{Chalupsky2013}%
  \BibitemOpen
\bibfield  {number} {  }\bibfield  {author} {\bibinfo {author} {\bibfnamefont
  {J.}~\bibnamefont {Chalupsk\'{y}}}, \bibinfo {author} {\bibfnamefont
  {T.}~\bibnamefont {Burian}}, \bibinfo {author} {\bibfnamefont
  {V.}~\bibnamefont {H\'{a}jkov\'{a}}}, \bibinfo {author} {\bibfnamefont
  {L.}~\bibnamefont {Juha}}, \bibinfo {author} {\bibfnamefont {T.}~\bibnamefont
  {Polcar}}, \bibinfo {author} {\bibfnamefont {J.}~\bibnamefont {Gaudin}},
  \bibinfo {author} {\bibfnamefont {M.}~\bibnamefont {Nagasono}}, \bibinfo
  {author} {\bibfnamefont {R.}~\bibnamefont {Sobierajski}}, \bibinfo {author}
  {\bibfnamefont {M.}~\bibnamefont {Yabashi}},\ and\ \bibinfo {author}
  {\bibfnamefont {J.}~\bibnamefont {Krzywinski}},\ }\href
  {https://doi.org/10.1364/OE.21.026363} {\bibfield  {journal} {\bibinfo
  {journal} {Opt. Express}\ }\textbf {\bibinfo {volume} {21}},\ \bibinfo
  {pages} {26363} (\bibinfo {year} {2013})}\BibitemShut {NoStop}%
\bibitem [{\citenamefont {Kujala}\ \emph {et~al.}(2020)\citenamefont {Kujala},
  \citenamefont {Freund}, \citenamefont {Liu}, \citenamefont {Koch},
  \citenamefont {Falk}, \citenamefont {Planas}, \citenamefont {Dietrich},
  \citenamefont {Laksman}, \citenamefont {Maltezopoulos}, \citenamefont
  {Risch}, \citenamefont {Dall’Antonia},\ and\ \citenamefont
  {Grünert}}]{Kujala2020}%
  \BibitemOpen
  \bibfield  {author} {\bibinfo {author} {\bibfnamefont {N.}~\bibnamefont
  {Kujala}}, \bibinfo {author} {\bibfnamefont {W.}~\bibnamefont {Freund}},
  \bibinfo {author} {\bibfnamefont {J.}~\bibnamefont {Liu}}, \bibinfo {author}
  {\bibfnamefont {A.}~\bibnamefont {Koch}}, \bibinfo {author} {\bibfnamefont
  {T.}~\bibnamefont {Falk}}, \bibinfo {author} {\bibfnamefont {M.}~\bibnamefont
  {Planas}}, \bibinfo {author} {\bibfnamefont {F.}~\bibnamefont {Dietrich}},
  \bibinfo {author} {\bibfnamefont {J.}~\bibnamefont {Laksman}}, \bibinfo
  {author} {\bibfnamefont {T.}~\bibnamefont {Maltezopoulos}}, \bibinfo {author}
  {\bibfnamefont {J.}~\bibnamefont {Risch}}, \bibinfo {author} {\bibfnamefont
  {F.}~\bibnamefont {Dall’Antonia}},\ and\ \bibinfo {author} {\bibfnamefont
  {J.}~\bibnamefont {Grünert}},\ }\href {https://doi.org/10.1063/5.0019935}
  {\bibfield  {journal} {\bibinfo  {journal} {Review of Scientific
  Instruments}\ }\textbf {\bibinfo {volume} {91}},\ \bibinfo {pages} {103101}
  (\bibinfo {year} {2020})},\ \Eprint
  {https://arxiv.org/abs/https://pubs.aip.org/aip/rsi/article-pdf/doi/10.1063/5.0019935/14876655/103101\_1\_online.pdf}
  {https://pubs.aip.org/aip/rsi/article-pdf/doi/10.1063/5.0019935/14876655/103101\_1\_online.pdf}
  \BibitemShut {NoStop}%
\bibitem [{\citenamefont {de~la Varga}\ \emph {et~al.}(2013)\citenamefont
  {de~la Varga}, \citenamefont {Velarde}, \citenamefont {de~Gaufridy},
  \citenamefont {Portillo}, \citenamefont {Cotelo}, \citenamefont {Barbas},
  \citenamefont {Gonz{\'a}lez},\ and\ \citenamefont {Zeitoun}}]{Varga2013}%
  \BibitemOpen
  \bibfield  {author} {\bibinfo {author} {\bibfnamefont {A.~G.}\ \bibnamefont
  {de~la Varga}}, \bibinfo {author} {\bibfnamefont {P.}~\bibnamefont
  {Velarde}}, \bibinfo {author} {\bibfnamefont {F.}~\bibnamefont
  {de~Gaufridy}}, \bibinfo {author} {\bibfnamefont {D.}~\bibnamefont
  {Portillo}}, \bibinfo {author} {\bibfnamefont {M.}~\bibnamefont {Cotelo}},
  \bibinfo {author} {\bibfnamefont {A.}~\bibnamefont {Barbas}}, \bibinfo
  {author} {\bibfnamefont {A.}~\bibnamefont {Gonz{\'a}lez}},\ and\ \bibinfo
  {author} {\bibfnamefont {P.}~\bibnamefont {Zeitoun}},\ }\href
  {https://doi.org/http://dx.doi.org/10.1016/j.hedp.2013.05.010} {\ \textbf
  {\bibinfo {volume} {9}},\ \bibinfo {pages} {542 } (\bibinfo {year}
  {2013})}\BibitemShut {NoStop}%
\bibitem [{\citenamefont {Murti}\ \emph {et~al.}(2008)\citenamefont {Murti},
  \citenamefont {Raju}, \citenamefont {Seetharami~Reddy}, \citenamefont
  {Koteswara~Rao}, \citenamefont {Mombaswala}, \citenamefont {Seshi~Reddy},
  \citenamefont {Lakshminarayana},\ and\ \citenamefont
  {Premachand}}]{Murti2008}%
  \BibitemOpen
  \bibfield  {author} {\bibinfo {author} {\bibfnamefont {M.~V.~R.}\
  \bibnamefont {Murti}}, \bibinfo {author} {\bibfnamefont {S.~S.}\ \bibnamefont
  {Raju}}, \bibinfo {author} {\bibfnamefont {B.}~\bibnamefont
  {Seetharami~Reddy}}, \bibinfo {author} {\bibfnamefont {V.}~\bibnamefont
  {Koteswara~Rao}}, \bibinfo {author} {\bibfnamefont {L.~S.}\ \bibnamefont
  {Mombaswala}}, \bibinfo {author} {\bibfnamefont {T.}~\bibnamefont
  {Seshi~Reddy}}, \bibinfo {author} {\bibfnamefont {S.}~\bibnamefont
  {Lakshminarayana}},\ and\ \bibinfo {author} {\bibfnamefont {K.}~\bibnamefont
  {Premachand}},\ }\href {https://doi.org/10.1007/s12043-008-0035-y} {\bibfield
   {journal} {\bibinfo  {journal} {Pramana}\ }\textbf {\bibinfo {volume}
  {70}},\ \bibinfo {pages} {747} (\bibinfo {year} {2008})}\BibitemShut
  {NoStop}%
\bibitem [{\citenamefont {Oura}\ \emph {et~al.}(2002)\citenamefont {Oura},
  \citenamefont {Yamaoka}, \citenamefont {Kawatsura}, \citenamefont {Takahiro},
  \citenamefont {Takeshima}, \citenamefont {Zou}, \citenamefont {Hutton},
  \citenamefont {Ito}, \citenamefont {Awaya}, \citenamefont {Terasawa},
  \citenamefont {Sekioka},\ and\ \citenamefont {Mukoyama}}]{Oura2002}%
  \BibitemOpen
  \bibfield  {author} {\bibinfo {author} {\bibfnamefont {M.}~\bibnamefont
  {Oura}}, \bibinfo {author} {\bibfnamefont {H.}~\bibnamefont {Yamaoka}},
  \bibinfo {author} {\bibfnamefont {K.}~\bibnamefont {Kawatsura}}, \bibinfo
  {author} {\bibfnamefont {K.}~\bibnamefont {Takahiro}}, \bibinfo {author}
  {\bibfnamefont {N.}~\bibnamefont {Takeshima}}, \bibinfo {author}
  {\bibfnamefont {Y.}~\bibnamefont {Zou}}, \bibinfo {author} {\bibfnamefont
  {R.}~\bibnamefont {Hutton}}, \bibinfo {author} {\bibfnamefont
  {S.}~\bibnamefont {Ito}}, \bibinfo {author} {\bibfnamefont {Y.}~\bibnamefont
  {Awaya}}, \bibinfo {author} {\bibfnamefont {M.}~\bibnamefont {Terasawa}},
  \bibinfo {author} {\bibfnamefont {T.}~\bibnamefont {Sekioka}},\ and\ \bibinfo
  {author} {\bibfnamefont {T.}~\bibnamefont {Mukoyama}},\ }\href
  {https://doi.org/10.1088/0953-4075/35/18/305} {\bibfield  {journal} {\bibinfo
   {journal} {Journal of Physics B: Atomic, Molecular and Optical Physics}\
  }\textbf {\bibinfo {volume} {35}},\ \bibinfo {pages} {3847} (\bibinfo {year}
  {2002})}\BibitemShut {NoStop}%
\end{thebibliography}

\end{document}